# General Model for Infrastructure Multi-channel Wireless LANs

Abdelsalam Amer and Fayez Gebali

Department of Electrical and Computer Engineering, University of Victoria
Victoria, BC, V8W 3P6, Canada

e-mails:*{aamer,fayez}*@ece.uvic.ca

## Abstract

In this paper we develop an integrated model for request mechanism and data transmission in multi-channel wireless local area networks. We calculated the performance parameters for single and multi-channel wireless networks when the channel is noisy. The proposed model is general it can be applied to different wireless networks such as IEEE802.11x, IEEE802.16, CDMA operated networks and Hiperlan\2.

## Index Terms

*Cross-layer model, request channels, data channels, error control, IEEE802.11x, IEEE802.16, Hiperlan\2, CDMA*

## 1. INTRODUCTION

Data integrity in wireless local area network is very important and providing an accurate and integrated model for request mechanism and data transmission is a challenge. The centralized wireless network where the Access Point (AP) or the Base Stations (BS) allocates the resources is considered in this paper. SSs issue requests in one of the logical channels. The BS receives the requests and issues grants to the SSs. The SSs once received the grant start sending their data on the data logical channels. Resources (logical channels) in wireless local area networks (WLANs) are limited and the demand for access to the medium leads to competition and collisions. Collisions reduce the throughput, increase delay and energy needed to transmit a packet. Thus, certain algorithms have to be developed in order to resolve the contention and get better Medium Access Control (MAC). The MAC frame has limited duration and is composed of several phases. Subscriber Stations (SSs) compete for access on the random access channels. Once SSs are granted the access the Base Station (BS) allocates data channels on the uplink and downlink phases for data transmission. Therefore, allocating extra requesting channels may reduce the collision probability but reduces the logical channels dedicated to uplink data transmission. On the other hand, reducing the number of requesting channels increases the collisions and as a result SSs encounter longer delay. A balance point between the number of requesting channels and data channels is a challenge. The physical layers of wireless standards are very similar and are based on the use of Orthogonal Frequency Division Multiplexing (OFDM). OFDM is used to combat the frequency selective fading and to randomize the burst errors caused by wideband fading channel.

In this paper we propose a cross-layer model for request mechanism and data transmission channels. We allocate two groups of channels in the MAC frame; requesting channels and data channels. The requesting channels are used by SSs to issue request for grants whereas the data channels are allocated by the BS to the SSs to send data in the uplink. We studied the requests performance and data performance while we vary the number of these channels. This model is a general model since it can be applied to different wireless standards. If we have only one access channel then it can be applied to IEEE802.11a [1]. If we consider the channels as frequency channels then it can be applied to WiMAX [2], and if the channels are time slots then it can be applied to Hiperlan\2 [3]. This paper is organized as follows; Section 2 presents the related work. Section 3 presents the proposed network model. Section 4 presets our results and comments. Conclusions are drawn in Section 5.

## 2. RELATED WORK

The medium access control in IEEE802.11x is distributed MAC protocol based on Carrier Sense Multiple Access with Collision Avoidance (CSMA/CA) while Hiperlan\2 is based on Time Division Duplexing TDD Time Division Multiple Access TDMA. The BS manages the SSs access to the MAC frame. In CDMA networks Cai et al. in [4] studied the performance of the CDMA random access system with linear minimum mean-squared error and MF receivers and the diversity combining in fading channels. In [5] Cooper et al. investigated the problem of random-access channel performance as it pertains to wide-band code-division multiple-access (W-CDMA) wireless systems. In [6] Zhao, studied DS-CDMA with slotted aloha random access protocols. In his work he distinguished between the two stages in transmission process, the access stage and the reception stage. Several cross-layer models have been proposed in WLANs [7]. Bouam in [8] proposed a cross-layer design in which IEEE802.11b MAC layer used knowledge of 802.11b physical layer state to manage the channel access. Alonso et al. proposed several models in cross-layer design and QoS support using Distributed Queuing Collision Avoidance DQCA. In his work, he proposed cross-layer resource management mechanisms for voice and data traffic that combine service differentiation and opportunistic transmission [9]. In other work he also proposed a smart scheduling algorithms that operate over a near optimum MAC protocol named Distributed Queuing collision avoidance and enhance its performance [10], [11], [12], [13]. B.Walke et al. in [14], studied the performance of Hiperlan\2. In his work he presented different models for physical and data link layer. However, he did not consider the cross layer modeling. Random access and collision reduction in Hiperlan$n$2 also been discussed in [15], [16], [17], [18]. In these papers, the random access channels are added based on the collision occurs in the previous MAC frame and they reduced if no access requests been issued. Also, the allocation of two slots in random access channels for each collided request reduces the MAC frame duration since the increase of random channels will affect other phases durations'.

Wireless channel is prone to errors due to noise and fading. Therefore error control protocol has to be applied to deliver safe data to the receiver. Automatic-repeat-request (ARQ) techniques are used to control transmission errors. Corrupted frames have to be retransmitted in whole or only the corrupted packets in the frame. Hui Li et al. in [19] presented selective repeat and request with partial bitmap. Despite the lower overhead, still the throughput is low. Atsushi proposed PRIME-ARQ [20] that improved the throughput but lacks the flexibility to be used in different wireless standards. A. Afonso in [21] proposed an algorithm for fast retransmission and adaptive rate scheme to reduce the delay, however, the scheme reserves some bandwidth which might be not used and hence the MAC utilization is effected. Other models have been proposed but they only considered one connection or the channel error was neglected [23], [24], [25].

## 3. NETWORK MODEL

In this section we show an example of a network model. Figure 1 shows using Time Division Duplexing (TDD) where time is broken down into frames and each frame has uplink and downlink phases. These phases use logical channels. The logical channels could be time slots in case of networks that use the TDMA (Hiperlan\2) as their medium access. They can be frequencies for networks that have frequency domain their medium access (WiMAX) or codes in CDMA network (3G). The uplink phase has two groups of channels, the uplink data channels and the request channels. The request channels are used when SS issue request whereas the data channels dedicated for SSs data. The downlink phase also has two groups of channels, the grant and the downlink data channels. The grant channels are used to inform the SS's with their allocated data channels whereas the downlink data channels are for the users data. A SS requests access using one of the $k$ request logical channels. The base station receives the requests and issues grants to the SS. Once the SSs receive their allocated grants, they start sending their data on one or more of the $L$ data logical channels. The balance point between the resources dedicated to the requesting channels $k$ and the data channels $L$ is a challenge. Increasing $k$ will reduce the request collision probability but affects the allocated data channels and result in degradation in the performance. On the other hand, reducing $k$ will increase the collision and as a result access delay is high.

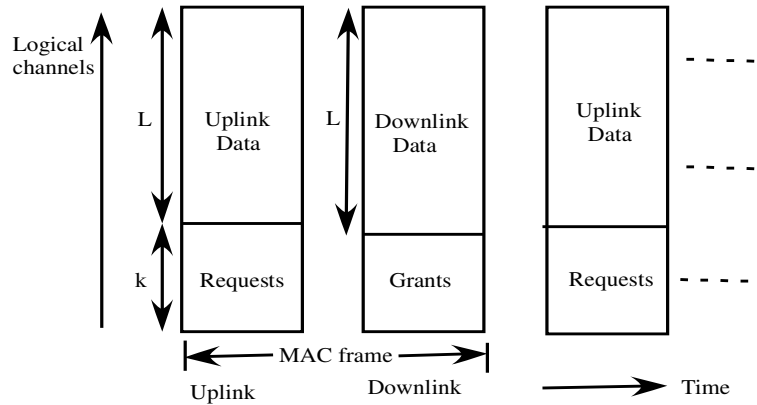

Fig. 1: Uplink and Downlink chart TDD

## A. Modeling channel error

Wireless channel could experience different channel fading. We consider digitized voice with $BER = 10^{-3}$ is an acceptable error rate because it is in general cannot be detected by the human ear. To maintain $BER = 10^{-3}$ with BPSK modulation with Rayleigh fading channel we need 24*dB* and it requires $SNR = 8dB$ for AWGN and 14*dB* for Rician channel as shown in Fig. 2a. Fig. 2b shows the *SNR* versus *BER* for different channels in 16-QAM modulation scheme. Since 16-QAM modulation scheme is higher than BPSK so it requires more *SNR* to get the same desired *BER* We considered different wireless channels (AWGN, Rayleigh fading channel and Rician channel) with different modulation schemes (BPSK and 16-QAM). BPSK is used as a fundamental mode to send control data in most of the wireless networks since it does not requires high SNR. The typical minimum SNR required for acceptable performance is 24*dB* [26]. These values will be used in our numerical simulations. We obtained these results using MATLAB.

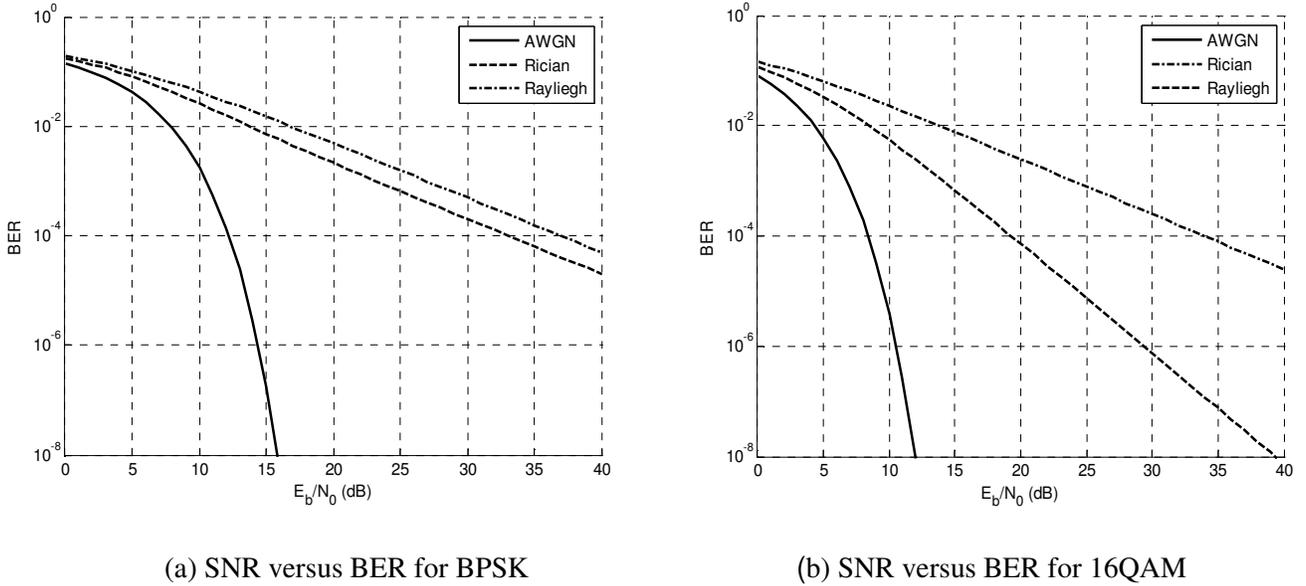

  (a) SNR versus BER for BPSK        (b) SNR versus BER for 16QAM

Fig. 2: SNR versus BER for different modulations and channels

## B. Model assumptions

We assume that *N* users try to request access on the random requesting channels. The number of request channels is assumed to be *k*. In order to analyze the system behaviour, some assumptions are made;
1) The probability that a user issues a request is *a*.
2) The probability a user chooses a particular reservation channel is $1/k$.
3) A collided user retransmits with probability *c*.
4) The traffic is calculated in one radio cell. No outside traffic is considered.
5) The average length of a packet is *b* bits.
6) The feedback channel is error free.
7) The sender will keep retransmitting a packet *n* attempts.
8) The time step is taken equal to the sum of transmission delay (time required to send a frame) and round trip delay (time required for frame propagation and reception of acknowledgment).
9) The forward channel has random noise and the probability that a bit will be received in error is *ε*, (*BER*).

Fig. 3 shows integrated model Markov chain state diagram for request and data with error control. The collided users adapts a constant probability backoff algorithm in which the collided users retransmit with a probability *c* [27].

The error control and MAC protocol states for requests and data are shown in Fig. 3 represented by *st*0 until *stn*. State $s_{ti}$ indicates that the SS is retransmitting the frame for the $i^{th}$ time whereas; state *st*0 indicates error-free transmission. A Subscriber Station (SS) that has data to send issues a request on one of the *k* random requesting channels. Contention may occur if two or more SSs choose the same requesting channel. A user could be in one of three states; *transmit* state, if a single request received or *collide* state, if two or more SSs issue a request on the same channel or *idle* state if there is no frame has been received. The SS will keep sending the packet if there is no acknowledgment is received (i.e the packet sent with an error probability *e*) *n* times. When a packet is correctly received the SS goes to *idle* state with probability *1-e*.
The error is calculated by;

$$e = 1-(1-\varepsilon)^b \qquad (1)$$

where *b* is the number of bits in a message. The probability that a user successfully accesses one of the free channels is given by;

$$x = (1-\frac{1}{k})^{N_{ave}-1} \qquad (2)$$

where $N_{ave}$ is the average number of active users;

$$N_{ave} = N(as_i + cs_c) \qquad (3)$$

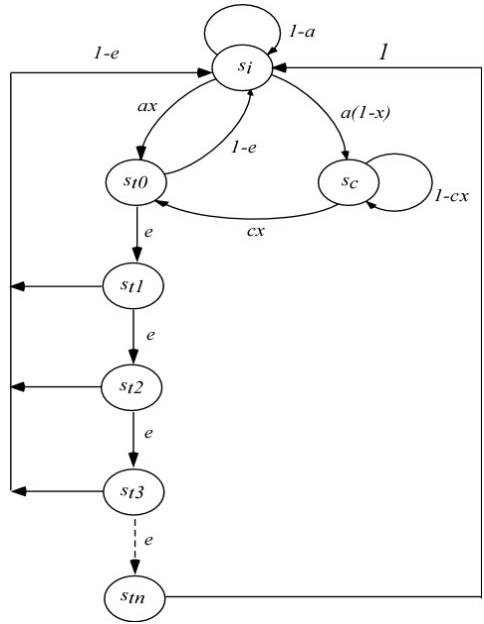

Fig. 3: Markov state diagram for requests and data channels

A discrete-time Markov chain is characterized by the transition matrix **P** which can be obtained from the state diagram and the state vector **s** [34]. The state vector s for the user is organized as follows;

$$s = [s_i \ s_c \ s_{t0} \ s_{t1} \ s_{t2} \ s_{t3} \ \text{------} \ s_{tn}]^t \qquad (4)$$

where $s_i$ is the probability that the user is in the *idle* state, $s_t$ is the probability that the user is in the *transmit* state and $s_c$ is the probability that the user is in the *collide* state.
At equilibrium, the distribution vector elements are obtained by solving the following two equations [28];

$$Ps = s \qquad (5)$$

$$\sum s_j = 1 \tag{6}$$

Where $j \in \{i, t0, t1, t2, \ldots tn, c\}$. From Eqs. (5) and (6) we can find the state vector elements at equilibrium [29]

$$s_i = \frac{1}{D}$$

$$s_c = \frac{a(1-x)}{D}$$

$$s_{tj} = \frac{B}{D} \sum_{j=0}^{n-1} e^j \tag{7}$$

Where $B = ax[1 + c(1-x)]$
and

$$D = 1 + B + eB + e^2 B + e^3 B + \ldots e^{n-1} B + a(1-x) \tag{8}$$

## C. Model performance

We study the performance of this model in this subsection.
The requests throughput is obtained from the following equation:

$$Th = \min(Ns_t, k) \tag{9}$$

The requests acceptance probability is defined as the ratio between the throughput and the offered load [28]:

$$p_a = \frac{Th}{Na} \tag{10}$$

The requests access delay is the average number of access attempts made by the SSs before they are successfully granted a channel. It is calculated by;

$$\tau = \sum_{i=0}^{\infty} i(1-p_a)^i p_a$$

$$= \frac{1-p_a}{p_a} \tag{11}$$

The requests average energy $E_a$ required to transmit a request successfully can be calculated as follows [30];

$$E_a = E_0 \sum_{i=0}^{\infty} (i+1)(1-P_a)^i P_a$$

$$= \frac{E_0}{P_a}$$

$$E_a[dB] = -10 \log(p_a) \tag{12}$$

where $E_0$ is the energy required to transmit a request once.
The Data channel throughput can be calculated by;

$$Th_{data} = \min\{L, Ns_t\} \tag{13}$$

Data channel acceptance probability can be calculated by:

$$P_{a(data)} = \frac{Th_{data}}{Na} \tag{14}$$

The data channel average delay can be calculated by;

$$\tau_{data} = \frac{1 - p_{a(data)}}{p_{a(data)}} \tag{15}$$

The data channel average energy can be calculated by;

$$E_{data}[dB] = -10 \log p_{a(data)} \tag{16}$$

## 4. RESULTS

In this section we will present our results. In the performance we assume that we have $N = 50$, retransmission probability $c=0.75$, the requesting channels is fixed at $k = 25$ when we study the impact of varying the data channels $L = \{1, 5, 10, 15\}$, $\varepsilon = 10^{-3}$. We fixed the data channels $L = 10$ when we study the impact of the requesting channels as we vary their values $k = \{10, 20, 25\}$. Fig. 4 shows the requests and data throughput. In Fig. 4a, requests throughput is shown and it is not effected by varying the number of data channels $L$. It is only affected by the number of requesting channels $k$. As the number of requesting channels increases the requests throughput increase. However, in the heavy the offered load the requesting throughput decreases as a result of collision since the collided users retransmit. In Fig. 4b the data throughput is presented. In the figure we can see that as we vary the number of data channels $L$ the throughput increases until it reaches the value of the requests throughput. For that reason, the two curves for $L=10$ and $15$ appears on one line, and that is always the case we study the performance of the data channels. That means all the granted users' requests' get their data transmitted. The worst data throughput is when we have only one data channel and that is a special case for IEEE802.11a.

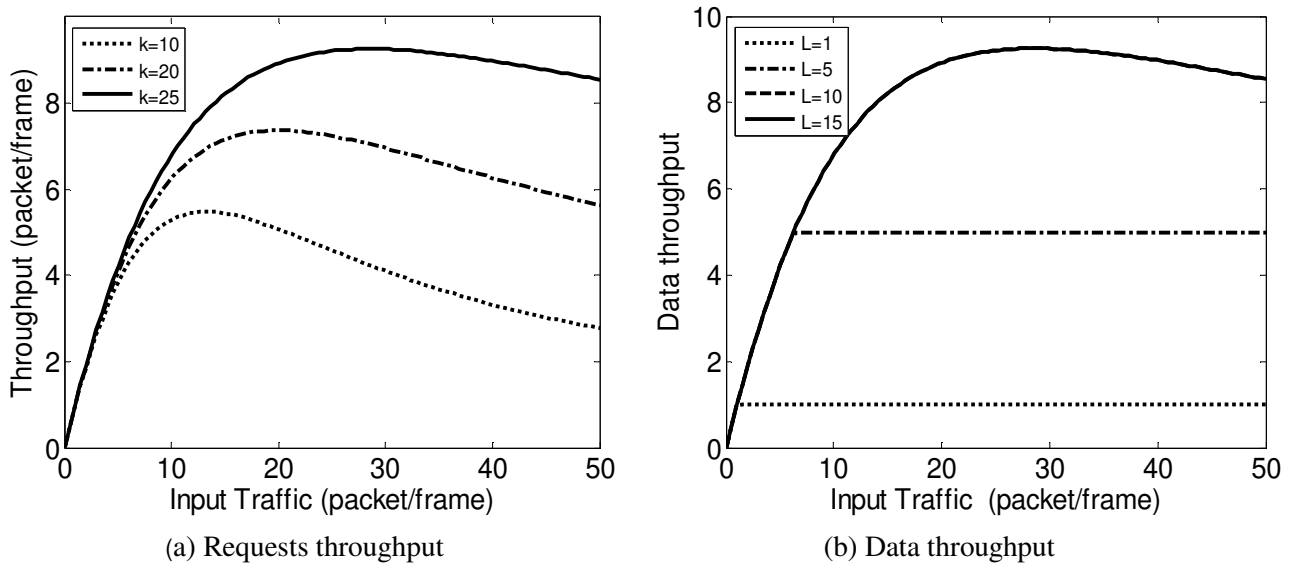

(a) Requests throughput  (b) Data throughput

Fig. 4: Requests and Data throughput

Fig. 5 shows the acceptance probability. The requests and data acceptance probability are at maximum value when we have a very small load and small number of users are requesting. However, as the load increase the acceptance probability decreases and this is normal since many users are competing for access and collision increases. The IEEE802.11a has the lowest acceptance probability since it only has one access channel and the collision is higher compared to other wireless standards. The requests and data acceptance probability are the same when we have many data channels and all successful requests get data channels to send their data. The acceptance requesting probability varies with number of requesting channels *k* and not affected by the number of data channels *L*

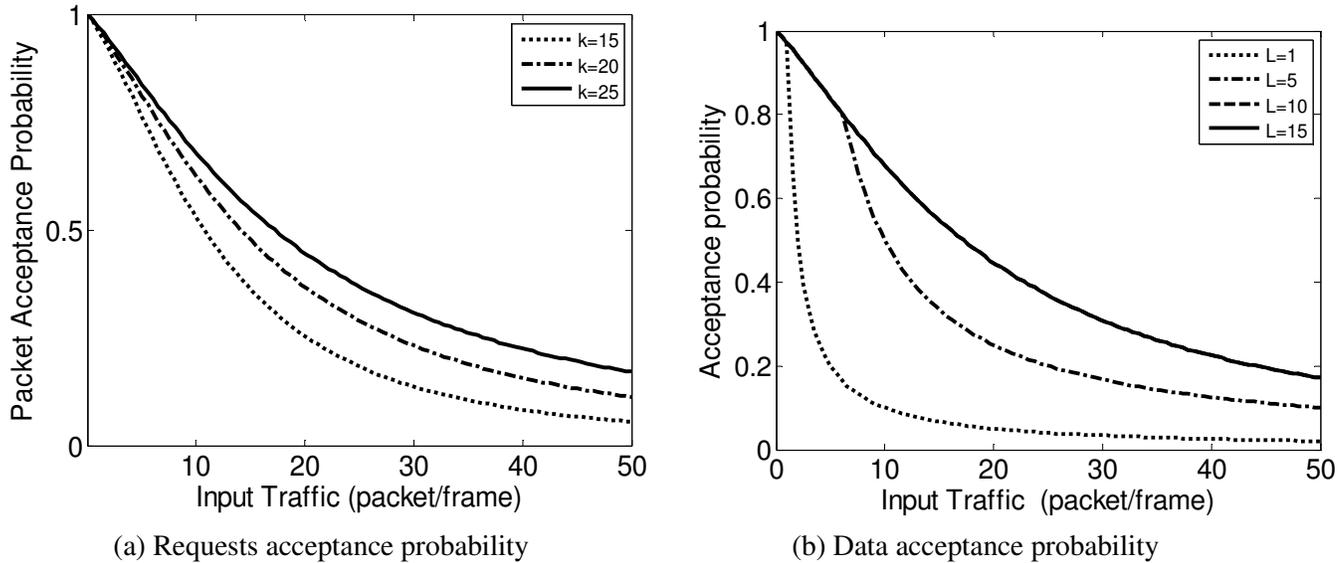

(a) Requests acceptance probability

(b) Data acceptance probability

Fig. 5: Requests and data acceptance probability

Fig. 6 shows our results for the average access delay. The one channel standard encounters the longest delay. The delay decreases as the number of data channels increases until the average requests access delay is similar to the average data delay. The requesting delay decreases as the number of requesting channels increases. However, increasing the requesting channels affects the MAC frame duration and as a result the data transmission delay affected.

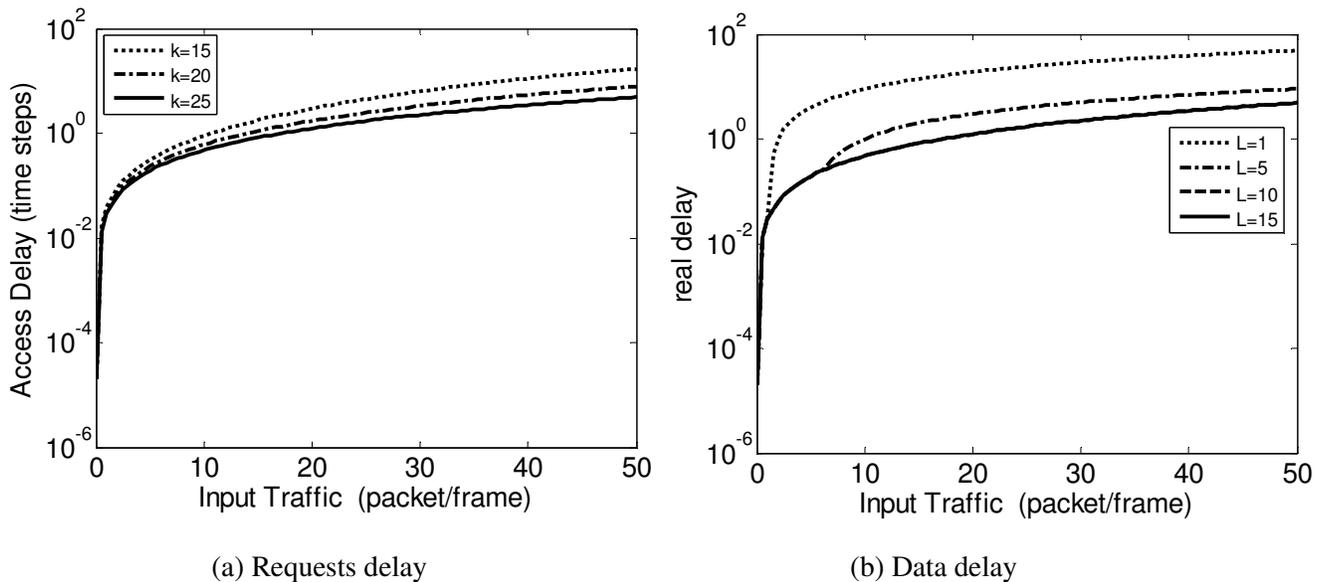

(a) Requests delay

(b) Data delay

Fig. 6: Requests and data delay

The average energy comparison is shown in Fig. 7. The average request energy is the same as the data average energy when the number of data channels is high. However, the average data energy is higher when we have less number of data channels as shown in Fig. 7b.

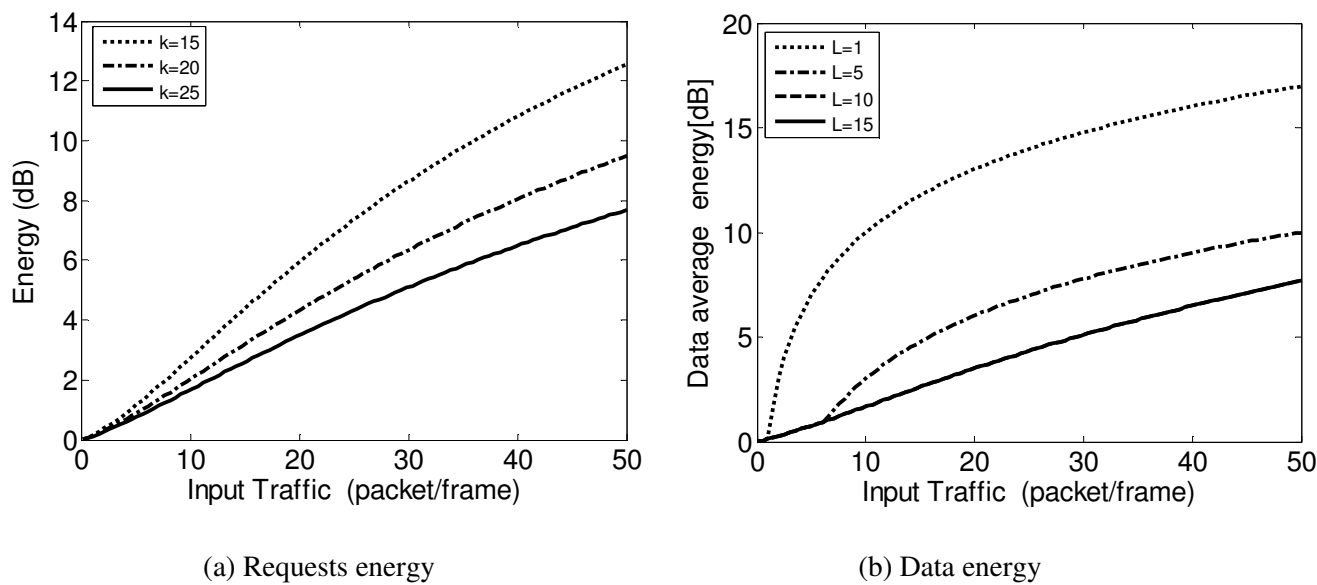

(a) Requests energy  (b) Data energy

Fig. 7: Requests and data energy

## 5. CONCLUSIONS

In this paper we developed a model for request mechanism and data transmission channels in wireless networks (4-way handshaking). We studied the impact of varying several parameters such as the requesting channels and the data channels. The requests performance in dependant only on the number of requesting channels whereas the data performance is depended only on the data channels. The requesting channels is independent of the data channels. As the number of requesting channels increases the performance of the requests improved and vise versa. Also, the data performance improved as the number of data channels increases. However, we can not allocate many number of requests channels since that is affects the data channels. So, the balance point between the number of requesting channels and data channels is a challenge. We noticed that the single channel wireless standard performs badly compared to multichannel standards.

International Journal of Computer Networks & Communications (IJCNC), Vol.2, No.3, May 2010